# Sort Race

HANTAO ZHANG, The University of Iowa
BAOLUO MENG, The University of Iowa
YIWEN LIANG, Wuhan University

Sorting is one of the oldest computing problems and is still very important in the age of big data. Various algorithms and implementation techniques have been proposed. In this study, we focus on comparison based, internal sorting algorithms. We created 12 data types of various sizes for experiments and tested extensively various implementations in a single setting. Using some effective techniques, we discovered that quicksort is adaptive to nearly sorted inputs and is still the best overall sorting algorithm. We also identified which techniques are effective in timsort, one of the most popular and efficient sorting method based on natural mergesort, and created our version of mergesort, which runs faster than timsort on nearly sorted instances. Our implementations of quicksort and mergesort are different from other implementations reported in all textbooks or research articles, faster than any version of the C library qsort functions, not only for randomly generated data, but also for various types of nearly sorted data. This experiment can help the user choose the best sorting algorithm for the hard sorting job at hand. This work provides a platform for anyone to test their own sorting algorithm against the best in the field.

• Sorting algorithms

## 1. INTRODUCTION

What is the beat sorting algorithm? The best answer to this question is perhaps "it depends". If you are asked to recommend a sorting algorithm for an unknown problem, what is your recommendation?

Sorting is perhaps the most studied computing problem. Various algorithms and implementation techniques have been used to demonstrate various algorithm design techniques. At *http://www.sorting-algorithms.com*, eight different sorting methods, that is, insertion sort, selection sort, bubble sort, shellsort, mergesort, heapsort, and two versions of quicksort, are animated on four types of initial instances: Random, Nearly Sorted, Reversed, and Few Unique. Their conclusion is that "there is no best sorting algorithm", and "the initial condition (input order and key distribution) affects performance as much as the algorithm choice." Apparently, this conclusion does not provide any clue to the aforementioned recommendation. From now on, we will use *Demo* as shorthand for *http://www.sorting-algorithms.com*.

In a decathlon competition, the winner is not necessarily the fastest runner or the highest jumper, but is determined by the combined performance in all events. If we choose the best implementation for each sorting algorithm and put them in a rigorous sort race, the winner can be decided by the combined performance in all tests. As in any relay competition, the fairest performance metric would be the total time taken by each method and we will use it in our sort race.

To make the sort race simple, we choose only comparison-based internal sorting algorithms. We will use instances of at least one million elements. The same instance and the same comparison function will be used by all participants in the same way, and all implementations will be compiled into one single program. Since selection sort and bubble sort perform poorly on any large instance, they are disqualified from the race. Quicksort should not be favored by giving it two places. Thus, we have five different sorting methods as participants in this race: insertion sort, shellsort [Knuth, D.E. 1997], mergesort [McIlroy, P. 1993], heapsort [Williams, J. W. J. 1964], and quicksort [Hoare, C.A.R. 1961].

In Table 1, we give the experimental results of the five sorting methods, plus the GNU C library qsort, Tim Peters' timsort [Peters, T. 2002], and David Musser's introsort [Musser, D. R, 1997], on four types of instances as used in *Demo*, i.e., random, reversely sorted, nearly sorted, and few unique inputs. All the methods are implemented in C and compiled as a single executable by gcc with optimization "-C3". The executable takes two parameters, the size and the type of the instance, generates one instance, and then run all the sorting methods on this instance. This way, the same set of instances is used for each method.

In Table 1, under "best time" is the fastest average time (in microseconds) over 100 instances (each has 2,000,000 elements) for each of the seven methods. Under each method, the ratio of its running time to the best time is given. That is, the actual running time of that method is the multiplication of the ratio and the best time. If the ratio is 1.00, then the best time is produced by that method. In row "Total avg. time"', the sum of four average times (in microseconds) from each data type is given, again in the format of ratio to the best time.

Table 1: The average running time of seven sorting methods on 4 types of inputs

| n=200000 | best time | insert | heap | shell | merge | quick | qsort | timsort | intro |
|---:|---:|---:|---:|---:|---:|---:|---:|---:|---:|
| random | 221114 | - | 3.57 | 1.75 | 1.18 | 1.02 | 1.21 | 1.17 | 1.00 |
| reversely sorted | 4728 | - | 113.78 | 26.34 | 1.00 | 1.78 | 19.74 | 1.07 | 7.67 |
| nearly sorted | 93185 | 1.62 | 5.73 | 1.92 | 1.00 | 1.54 | 1.50 | 1.11 | 1.30 |
| few unique | 128097 | - | 5.88 | 2.39 | 1.64 | 1.00 | 1.73 | 1.61 | 1.26 |
| Total avg. time | | - | 5.18 | 1.98 | 1.14 | 1.00 | 1.42 | 1.13 | 1.21 |

Among the five sorting methods, it is clear from the table that quicksort is the best overall method according to the best average time. The second best method is mergesort, which has the best performance for reversely sorted and nearly sorted instance. The last three columns give the performance of two different mergesorts and one introspective sort, as a reference to show that our results are very competitive: qsort, the sorting function in the GNU C Standard Library, is 42% slower than our quicksort; timsort [Peters, T., 2002] is the second best performer among all eight methods; and introsort [Musser, D.R, 1997] is the fourth best performer. These three sorting methods are known as the best generic sorting methods and have been used in C, C++, and Python libraries.

Why the conclusion from our experiment is different from the one in *Demo*? Apparently there are two reasons: (1) size matters, and (2) implementation matters.

From action movies, we know some people run faster than trains, of course, in short races. However, it is wrong to conclude that human are faster than trains. In *Demo*, up to 50 elements are used to illustrate the idea of each sorting method, and this size is too small to demonstrate the strength of each method. For example, the overhead for choosing a good pivot in quicksort only pays off with a large size array. Moreover, it is very important to know that a sorting method works well or not for very large size problems. Every experienced programmer knows that different implementations of the same algorithm give different performances. For Olympics, each country sends the best athletes to compete. For sort race, we should choose the best implementation possible for each method.

When talking about the best sorting implementation, we cannot ignore the existence of the qsort() function in the C Standard Library. Overtime, qsort() tries to represent the best sorting function and there are at least 14 different implementations of qsort() developed. Nowadays, the most popular qsort() is the GNU project's libc implementation which powers most GNU/Linux distributions and several other operating systems. This qsort is interesting in that it shuns quicksort in favor of mergesort, due to the popularity of timsort [Peters, T. 2002]. Tim Peters created timsort based on Peter McIlroy's idea of natural mergesort [McIlroy, P. 1993], which consists of two processes: in process 1, sorted subarrays (called *runs*) are identified from the input; in process 2, two runs are merged into one run, until only one run remains. The two processes can be interleaved. There are several implementations based on natural mergesort, such as timsort [Peters, T. 2002], the GNU qsort, and neatsort [La Rocca *et al.* 2014]. Different techniques are used in these implementations. As the first goal of this article, we will try to identify which techniques are more effective and which are less effective through extensive experiments. From now on, by mergesort, we mean natural mergesort [McIlroy, P. 1993].

Before the GNU qsort, the most popular qsort is the BSD qsort, which is a version of quicksort based on Bentley and McIlroy's qsort function [Bentley and McIlroy, 1993]. A significant deviation from Bentley and McIlroy's qsort is that the BSD qsort resorts to insertion sort, not only for small arrays, but also whenever a partitioning round is completed without having moved any elements. The latter is intended to capture nearly sorted inputs, since insertion sort will handle these cases efficiently. However, this change may backfire. We found by accident that the BSD qsort performs very poorly on a set of instances. David Musser's introsort, which is a hybrid quicksort with insertsort and heapsort [Musser, D.R., 1997], was another attempt to make quicksort adaptive to nearly sorted inputs and was the default sorting function in C++. We found that introsort is not effective for several contrived classes of inputs in our platform. By extensive experiments, we will try to identify which techniques used in the past [Sedgewick, R. 1978] are effective and what new techniques can be used so that quicksort can handle better nearly sorted inputs and has better overall performances over all kinds of inputs. This is the second goal of this article.

The third goal of this article is to establish a software platform available in the public domain so that anyone can test his own sorting algorithm against other best implementations. Currently, the platform, called *sortrace*, is a C program which provides 12 types of inputs and over 20 implementations of sorting algorithms. Each type can take another parameter to decide its subtype. For instance, we have tested over a dozen of different sequences in the shellsort algorithm to see which one works best. We hope that anyone who claims to have found a better sorting algorithm can use this platform to sustain his claim by going through the sort race.

Before we select the best implementation for quicksort and mergesort, let us see what types of inputs are available in the platform.

## 2.  TWELVE CLASSES OF INPUTS

For the first five classes of inputs, each array is randomly generated and each array item is one of the following five types: long integer, double, lists of integers of 64, 256, and 1024 bytes, respectively. When the size of array is under 4,000,000, the best performer is introsort on random integers and doubles. For sizes larger than 4,000,000, quicksort is the best performer for all five random types.

For the next seven classes, besides the list length $n$, it takes another integer $k$, where $1 \leq k \leq n$. Before we present these seven classes, let us first review some concepts.

Given an array $A = (a_1, a_2, ..., a_n)$, let $rank(a_i)$ be the final position of $a_i$ in the sorted array. Many researchers prefer to use the number of inversions, $Inv(A)$, and the number of ascending runs in the list, $Run(A)$, to measure the presortedness of $A$:

$$Inv(A) = |\{ (i, j) : 1 \leq i < j \leq n, a_i > a_j\}|$$
$$Run(A) = |\{ i : 1 \leq i < n, a_i > a_{i+1} \}| + 1$$

When $A$ is reversely sorted, the number of inversions is maximal ($Inv(A) = n(n–1)/2$) and the number of runs is also maximal ($Run(A) = n$). In this case, an $O(n)$ operation will make $A$ sorted and the presortedness of $A$ is not reflected in $Inv(A)$ and $Run(A)$. Hence, we prefer another measure over

$Inv(A)$ and $Run(A)$. For instance, we are interested in the minimal number $k$ such that $A$ can be broken into $k$ sublarrays and each subarray is monotonic (i.e., either sorted or reversely sorted).

$$Mono(A) = min\{\ k : A = w_1 w_2 \dots w_k,\ w_i \text{ is either sorted or reversely sorted, } 1 \leq i \leq k.\ \}$$

Intuitively, if $Mono(A) = k$, then $A$ is the concatenation of $k$ monotonic lists (either sorted or reversely sorted). For instance, if $A$ is sorted or reversely sorted, then $Mono(A) = 1$. If $A$ is an "organ-pipe" array, $Mono(A) = 2$.

Here are the five classes of inputs with an additional parameter $k$:

- **$k$-limited**: A simple way to create Few Unique instances is to restrict elements in a certain range. An array of integers is said to be *k-limited* if the number of distinct elements is bound by $2^k$. To create a $k$-limited array, we at first create randomly an array of integers at most $k$ bits, and then apply the function $f(x) = (1+x)^{60/k}$ to every array element if $k < 60$. The last step is called "blow up", which will make the array hard for sorting methods such as counting sort or radix sort. When $k = 0$, we assume that all elements of the array are the same, i.e., 0. The Few Unique instance used in *Demo* can be mapped to a 2-limited array of 50 elements. If we keep the same ratio of $2^2/50 = 2^k/2{,}000{,}000$, we get $k = 12$. For the Few Unique instance in Table 1, we used 12-limited arrays. When $k \geq 32$, it is the same as randomized integers. The best sorting method for the $k$-limited is the 3-way quicksort [Bentley and McIlroy, 1993].

- **$k$-equal teeth**: Divide $n$ positions into $k$ equal sections. Each section is a sorted subarray from 1 to $n/k$. Then we blow up the array by applying function $f(x) = x^{60/t}$ to each array element, where $t = \log_2(n/k)$. Such an array is called *k-equal teeth* (or *k-equal saw teeth*). Obviously, if $A$ is a $k$-equal teeth, $Run(A) = Mono(A) = k$. If $k = 1$, then $A$ is sorted. Among all comparison-based sorting algorithms, mergesort is the best one for this class of instances.

- **$k$-even teeth**: A *k-even teeth* array can be constructed from $k$-equal teeth: Simply reverse each odd numbered sections. When $k = 1$, the array is reversely sorted. For $k = 2$, it is an "organ-pipe" array. Obviously, for $k \leq n/3$, if $A$ is a $k$-even teeth, $Mono(A) = k$. When $k = n/2$, before applying function $f(x)$, the array is a 2-limited list starting with (2,1,1,2,2,1,1,2, ...), and $Mono(A) = n/4+1$, because we may group the list as (2,1,1), (2,2,1,1), (2,2,1,1), …. The best sorting method for this class of instances is mergesort.

- **$k$-sharp teeth**: A *k-sharp teeth* array is constructed as follows: Initially, the array contains a sorted list from 1 to $n$. We divide the array into $k$ equal sections and reverse the odd numbered sections. Then we apply function $f(x) = x^{60/t}$ to each array element, where $t = \log_2(n)$. When $k = 1$, the array is reversely sorted. If $A$ is a $k$-sharp teeth, $Mono(A) = k$. The best sorting method for this class of instances is mergesort. We like to point out that the BSD qsort performs very poorly on this type of inputs. For instance, when $n = 2{,}000{,}000$ and $k = 8$, mergsort takes 0.01 sec., Bentley & McIlroy's function takes 0.1 sec., and the BSD qsort takes 279.4 sec. That might explain why quicksort was shunned from the C library.

- **$k$-shuffled teeth**: A *k-shuffled teeth* array is constructed as follows: The array is first initialized with a $k$-sharp saw-teeth, and then all $k$ sections are shuffled into one (preserving the relative order of the elements in each section).The concept of $k$-shuffled teeth lists is related to the SMS measure in [Moffat *et al.* 1996], the least value $k$ such that the input list can be formed by interleaving $k$ monotone (either descending or ascending) sequences. Like the class of $k$-sharp teeth, mergesort is the best for this class when $k < n/10$; when $k > n/10$, quicksort is the best.

- **k-distance:** An array $A = (a_1, a_2, \ldots, a_n)$ is said to be *k-distance* if $|rank(a_i) - i| \leq k$ for all $1 \leq i \leq n$. That is, for each $a_i$ in $A$, the distance from its initial position to its final position is bound by $k$. To create a $k$-distance list, we first construct a sorted list from 1 to $n$, divide the list into $n/(k+1)$ sections, and shuffle each section of $k+1$ elements randomly. Finally, we blow up the array by applying function $f(x) = x^{60/t}$ to each array element, where $t = \log_2(n)$. If $k = n - 1$, the result is a random list. The Nearly Sorted instance used in *Demo* appears to be a 1-distance array of 50 elements. If we keep the same ratio of 1/50, we would use a 40000-distance array of 2,000,000 elements. Instead, we used 256-distance lists in Table 1. Indeed, the best performer for this class on 2,000,000 elements is insert sort if $k \leq 32$; for $k > 32$, mergesort is the best. Shaker sort, a bidirectional bubble sort [Black and Bockholt, 2009], is about twice slower than insertion sort for this class of inputs.

- **k-exchange**: An array $A$ is said to be a *k-exchange list* if it becomes sorted after no more than exchanges of $k$ pairs elements in $A$. In this case, $|\{a_i : rank(a_i) \neq i\}| \leq 2k$. To create a $k$-exchange array, we first create a sorted list from 1 to $n$, apply function $f(x) = x^{\log(n)}$, and then repeat the following operation $k$ times: randomly choose two positions to exchange them. The model of *k-exchange* arrays is related to the *Rem* measure (the minimal number of items that must be removed to leave a sorted sequence) in [Moffat *et al*. 1996]. It can be shown that if $A$ is a *k-exchange* array, then $Mono(A) \leq 2k+1$. Like the class of *k-shuffled* teeth, mergesort is the best for this class when $k \leq n$; when $k > n$, quicksort is the best.

The last six classes contain some kind of presortedness and can be called *nearly sorted* with proper $k$. We will present experimental results of various sorting methods on these instances in later sections. Roughly speaking, quicksort is the best method for the first six classes of inputs: five random arrays plus *k*-limited arrays. Mergesort is the best method for the remaining six classes of inputs, with the exception of *k-distance* arrays when $k \leq 32$ ($n = 2,000,000$); in this case mergesort is slightly slower than insertion sort. Both *k*-distance and *k*-exchange arrays model nearly sorted arrays by several researchers [Cook *et al*. 1980]. Our experimental results show clearly that mergesort is the best choice for nearly sorted arrays. The second best is quicksort, which is much better than the splaysort method reported in [Moffat *et al*. 1996]. Whenever we use the above 12 classes of instances, we will always include the best performer's result for each class so that the ratios of each method's run time to the best time for that class will be kept the same.

### 3. SELECT BEST MERGESORT

The first choice for selecting the best implementation of mergesort is perhaps timsort [Peters, T. 2002]. Timsort is a highly optimized implementation of mergesort with at least the following techniques:

(a) When scanning the input, if it finds a reversely sorted subsequence, it will reverse the subsequence to create a longer run. If the stable property is required, either the reversely sorted subsequence cannot contain equal elements (as done in timsort) or subarrays of equal elements are reversed twice (as done in our implementation).
(b) If a run in the input is shorter than the minimal length for a run, insertion sort is called to create a longer run.
(c) The minimal length can be preset, or adjusted so that if each run keeps the minimal length, then the number of runs is (or slightly less than) a power of two.
(d) Binary search is used in the insertion sort to locate the position of the element to be inserted in the sorted list.
(e) Runs are inserted in a stack. Let X, Y, Z be the sizes of the top three runs in the stack, then it maintains that X > Y+Z and Y > Z. If X < Y+Z, then Y is merged with the smaller of X and Z until the two conditions are satisfied.

(f) Merge is always done among two consecutive sorted subarrays, say A and B. Any element in A no bigger than the first element of B is removed from the merge process. Similarly, any element of B bigger than the last element of A is also removed from the merge process. Only the shorter list of A or B is copied into a temporary memory.
(g) Merge goes into the galloping mode if A (or B) wins for a consecutive number of times. In the galloping mode, the binary search is used to locate the position in A where the current element in B should go.

**Example**: Suppose the input $A = (1, 15, 18, 19, 20, 16, 11, 10, 9, 8, 7, 6, 5, 4, 3, 2)$. $Run(A) = 12$ and $Mono(A) = 2$, because $A$ consists of one sorted subarray from 1 to 20 (a run) and one reversely sorted subarray from 16 to 2 (a reversed run). Timsort will reverse the second subarray and merge two runs into one. During the merge process, the first number in $A$, i.e., 1, will be skipped because it is smaller than 2, the first number of the second run. Then 15 through 20 in the first run will be copied into a temporary memory. Once 1, 2, 3, 4 are moved into the final position, since the second run won three times in a row, it goes into the galloping mode: Using binary search, it finds that 15 falls in between 11 and 16 of the second run. So 5 through 11 will be moved to the final position without comparing them one by one against 15.

We do not have information on how the GNU qsort is implemented. For neatsort [La Rocca *et al.* 2014], (a) and (f) are used; a heuristic different from (e) is used to select two runs merging into one. We have implemented our own mergesort and have tested (a) – (g) except (e). We found that (a), (f) and (g) have great impact to the performance of mergesort, so all of our implementations will use them. We also found that (b) provides a little performance improvement for most cases. However, we found that (c) has almost no impact. The motivation behind (c) is to avoid the merge of a very long run with a very short run. If the sizes of each run cannot be the same, the number of runs can hardly be an expected power of two. For (d), we found that using binary search in insertion sort, the impact to the running time is marginal, but it reduces the number of comparisons noticeably. When expensive comparison functions are used, we still do not see the gain on running time. Instead of (e), we have tested two new strategies to select runs to merge:

($e_1$) Merge of runs takes multiple rounds: in each round, all runs are paired and merged into one.
($e_2$) Maintain a list of runs such that the length of each run is not bigger than the half of the length of its predecessor. If the condition is not true, merge this run with its predecessor. Such a list is called "half-down list".

It turns out that the heuristic ($e_1$) is weaker than the other heuristics, but none of (e), ($e_2$) or neatsort's heuristic makes a significant impact to the overall performance.

In Table 2, we give the performance of either mergesort implementations for the 12 classes of inputs given in the previous section. The either mergesort implementations are the GNU qsort, timsort, neatsort, and five variants of our mergesort, which differ by the used techniques:
- mer2: uses (b), (f), (g), ($e_1$)
- mer3: uses (a), (f), (g), ($e_1$)
- mer4: uses (a), (b), (f), (g), ($e_1$)
- mer5: uses (a), (f), (g), ($e_2$)
- mer6: uses (a), (b), (f), (g), ($e_2$)

In fact, mer2 is not natural mergesort, as runs are not discovered from the input but created by insertion sort. The code of mer6 is given in the appendix for reference.

The value of $k$ for the last seven $k$-classes of inputs is $2^i$, where $i = 0, 1, …, 8$. In the column "best", the best average running time of 100 executions, in microseconds, is given. It may be produced by other implementations not listed in the table. The running times of each method are given as a ratio to the best average time. If it is 1.00, then this is the best performer among all sorting methods in our experiment. The column "qsort" gives the running time of the GNU qsort. The columns mer2–mer6 give the running times of our five versions of mergesort.

Table 2: Performances of 8 mergesort implementations for 12 classes of inputs

| n=200000 | best | qsort | mer2 | mer3 | mer4 | mer5 | mer6 | Tim | neat |
|---|---|---|---|---|---|---|---|---|---|
| random 16-list | 764548 | 1.43 | 1.74 | 1.80 | 1.77 | 1.32 | 1.27 | 1.21 | 1.80 |
| random 64-list | 1279070 | 1.25 | 1.84 | 1.99 | 1.89 | 1.26 | 1.25 | 1.19 | 2.00 |
| random 256-list | 1530270 | 1.24 | 1.81 | 1.92 | 1.83 | 1.20 | 1.19 | 1.16 | 1.88 |
| random double | 249758 | 1.14 | 1.07 | 1.12 | 1.08 | 1.07 | 1.01 | 1.11 | 1.15 |
| random integer | 231084 | 1.17 | 1.11 | 1.16 | 1.12 | 1.12 | 1.06 | 1.13 | 1.17 |
| k-limited | 142186 | 1.53 | 1.30 | 1.39 | 1.30 | 1.37 | 1.26 | 1.35 | 1.51 |
| k-equal teeth | 25180 | 3.24 | 1.38 | 1.03 | 1.05 | 1.00 | 1.03 | 1.10 | 1.20 |
| k-even teeth | 24712 | 3.56 | 2.90 | 1.04 | 1.19 | 1.00 | 1.14 | 1.11 | 1.21 |
| k-sharp teeth | 13846 | 6.14 | 4.47 | 1.15 | 1.34 | 1.00 | 1.26 | 1.22 | 1.61 |
| k-shuffled teeth | 96920 | 1.54 | 1.04 | 1.14 | 1.00 | 1.16 | 1.01 | 1.07 | 1.49 |
| k-distance | 51051 | 2.15 | 1.00 | 1.43 | 1.05 | 1.49 | 1.10 | 1.27 | 1.38 |
| k-exchange | 10075 | 8.55 | 1.78 | 1.21 | 1.24 | 1.00 | 1.09 | 1.11 | 3.28 |
| **Average (12)** | | 1.24 | 1.56 | 1.64 | 1.57 | 1.11 | 1.09 | 1.07 | 1.65 |
| **Average (8)** | | 1.65 | 1.18 | 1.10 | 1.03 | 1.08 | 1.00 | 1.07 | 1.23 |

The last row gives the average running times of the last 8 classes of inputs (not the average of ratios). The second last row gives the average time of all 12 classes. Mer6 is the best performer of the last 8 classes of inputs; mer4 is the second best (3% slower) and timsort is the third (7% slower). The GNU qsort is the worst performer in the table (65% slower).

When all the 12 classes are considered, timsort is the best among all mergesort implementations in our experiment, and is only 7% slower than the best quicksort implementation. Mer6 is the second best (2% slower than timsort). The advantage of timsort over mer6 is shown on the first three randomized classes where expensive comparison functions are used; for random integers or doubles, mer6 is still faster than timsort, just like introsort is faster than quicksort for these two classes. The idea of (d) cannot reduce the running time of mer6 while it can reduce the number of comparisons. This may suggest that (e) is better than ($e_1$) and ($e_2$) for large size lists of heavy items. For all the 12 classes, the worst performers are mer3, mer4, and mer2, which are 64%, 57%, 56%, respectively, slower than the best. This shows that ($e_1$) is certainly weaker than ($e_2$).

For the three contrived classes, *k-equal teeth*, *k-even teeth*, and *k-sharp teeth*, mer5 is the best performer among all implementations, while the GNU qsort is at least three times slower than mer5. For the class of *k-exchange*, qsort is almost eight times slower than mer5.

## 4. SELECT BEST QUICKSORT

As the most popular sorting algorithm, quicksort [Hoare, C.A.R. 1961] has been extensively studied by many researchers and various techniques have been proposed to implement it. To select the best implementation of quicksort, the first consideration will be Bentley and McIlory's qsort function [Bentley and McIlroy, 1993]. Let $n$ be the input array size, the following techniques are in Bentley and McIlory's qsort function:

(a) If $n < \alpha$, call insertion sort. Bentley and McIlroy use $\alpha = 7$.
(b) If $n < \beta$, use the median of the first, middle and the last elements as pivot for splitting the array. Bentley and McIlory use $\beta = 40$.
(c) If $n \geq \beta$, use the pseudo-median of 9 elements (taken at every $n/8$th position plus the first element) as pivot.
(d) During the splitting process, all elements equal to the pivot are placed either at the beginning or end of the array. When the splitting is done, these elements equal to the pivot are moved in the middle of the array and are spared from the recursive calls.
(e) Do not identify the position of the pivot in the array. Bentley and McIlroy use this idea for all eight byte elements.

The use of (c) is to guarantee that a good pivot element is chosen, to avoid the quadratic worst-case time complexity. The use of (a) is to avoid the overhead of choosing a better pivot. The idea behind (d) is called 3-*way splitting*, and quicksort using (d) is called the 3-way quicksort. Evidently, when no repeated elements are present, 3-way splitting is slower than the classic 2-way splitting [Cormen *et al.*, 2009]. The idea of (e) is to avoid swapping the pivot with other elements so that the presortedness of the input is not disturbed.

For the 3-way splitting, at the end of splitting, the array is divided into three parts: the first part contains elements less than the pivot; the middle part contains elements equal to the pivot and the last part contains elements greater than the pivot. Recursive calls apply to the first and the last parts. The 2-way splitting in any standard textbook [Cormen *et al.*, 2009] also splits the array into three parts: the elements less than the pivot, the pivot, and the elements no less than the pivot. Using (e) without (d), the array is split into two parts: the elements no greater than the pivot and the elements no less than the pivot.

The BSD qsort function in the C library has some deviation from Bentley and McIlroy's qsort: the BSD qsort also use insertion sort instead of recursive calls of qsort, whenever a splitting round is completed without having moved any elements. This idea is intended to capture nearly sorted inputs and may perform poorly for certain inputs. For instance, if an array contains $2m+1$ distinct integers, the first $m$ integers are the reversed 1 through $m$, the $(m+1)$th number is $m+1$, and the last $m$ integers are the reversed $m+2$ through $2m+1$, then the BSD qsort will use insertion sort to sort two reversed lists of $m$ elements. We have tested the BSD qsort on the 12 classes of inputs and its performance is very poor for the class of $k$-sharp teeth. Besides the ideas used by Bentley and McIlroy, we also have tested the following two ideas:

(f) If $n \geq \alpha$, test if the array is already sorted; if yes, exit.
(g) At first, use the 2-way splitting. If the array contains more than $\gamma = 2$ elements equal to the pivot, use the 3-way splitting in the recursive calls.

The use of (f) is to avoid wasting time on already sorted subarrays, which occur often in nearly sorted inputs. The use of (g) is a typical example of hybrid algorithms: The program starts with the 2-way quicksort; when it identifies repeated elements, it turns to the 3-way quicksort.

Combing (e) and (f), reversely sorted lists are no longer a problem for quicksort: Using the median of the input as pivot, the splitting will reverse the list and the two recursive calls will finish with (f). In other words, quicksort becomes adaptive as it takes O($n$) time on sorted or reversely sorted lists. It also works well for some other troublesome instances including organ-pipe lists.

Table 3: Performances of quicksort and mergesort implementations for 12 classes of inputs

| n=200000 | best | B&M | 3-way | 2-way | hyb1 | hyb2 | hyb3 | hyb4 | intro | mer6 | Tim | qsort |
|---|---|---|---|---|---|---|---|---|---|---|---|---|
| random 16-list | 764548 | 1.11 | 1.12 | 1.28 | 1.10 | 1.00 | 1.08 | 1.00 | 1.04 | 1.27 | 1.21 | 1.43 |
| random 64-list | 1279070 | 1.06 | 1.05 | 1.02 | 1.03 | 1.01 | 1.03 | 1.00 | 1.10 | 1.25 | 1.19 | 1.25 |
| random 256-list | 1530270 | 1.13 | 1.09 | 1.04 | 1.04 | 1.00 | 1.04 | 1.03 | 1.12 | 1.20 | 1.14 | 1.24 |
| random double | 249758 | 1.17 | 1.13 | 1.05 | 1.04 | 1.05 | 1.02 | 1.06 | 1.00 | 1.01 | 1.11 | 1.14 |
| random integer | 231084 | 1.18 | 1.11 | 1.03 | 1.03 | 1.05 | 1.02 | 1.05 | 1.00 | 1.06 | 1.13 | 1.17 |
| k-limited | 142186 | 1.13 | 1.07 | 1.10 | 1.03 | 1.04 | 1.00 | 1.04 | 1.22 | 1.26 | 1.35 | 1.53 |
| k-equal teeth | 25180 | 5.54 | 4.71 | 4.93 | 4.67 | 4.36 | 4.17 | 4.66 | 7.13 | 1.03 | 1.10 | 3.24 |
| k-even teeth | 24712 | 5.72 | 4.84 | 4.97 | 4.82 | 4.52 | 4.28 | 4.81 | 7.55 | 1.14 | 1.11 | 3.56 |
| k-sharp teeth | 13846 | 6.14 | 2.42 | 1.94 | 2.28 | 2.12 | 1.85 | 2.04 | 10.7 | 1.26 | 1.22 | 6.14 |
| k-shuffled teeth | 96920 | 2.37 | 2.09 | 1.95 | 1.96 | 1.97 | 1.88 | 1.98 | 1.88 | 1.01 | 1.07 | 1.54 |
| k-distance | 51051 | 2.53 | 2.41 | 2.04 | 2.09 | 2.17 | 1.85 | 2.21 | 1.59 | 1.10 | 1.27 | 2.15 |
| k-exchange | 10075 | 9.52 | 2.95 | 2.44 | 2.69 | 2.69 | 2.18 | 2.46 | 7.38 | 1.09 | 1.11 | 8.55 |
| **Average (12)** | 400872 | 1.23 | 1.06 | 1.04 | 1.02 | 1.00 | 1.01 | 1.01 | 1.11 | 1.09 | 1.08 | 1.24 |
| **Average (8)** | 82566 | 1.93 | 1.56 | 1.48 | 1.47 | 1.46 | 1.37 | 1.49 | 1.82 | 1.00 | 1.07 | 1.65 |

In Table 3, we provide the performances of seven implementations of quicksort plus David Musser's introsort [Musser, D.R., 1997], on the same set of inputs in Table 2. For reference, we also copied the results of three mergesorts from Table 2: mer6, Tim and the GNU qsort. The seven quicksort implementations are: (1) B&M, Bentley and McIlroy's qsort function. (2) 3-way: the 3-way quicksort uses the ideas of (a) to (f), and is almost identical to B&M except (f), i.e., test if the array is sorted as a preprocessing step. (3) 2-way: the 2-way quicksort using the ideas of (a), (b), (c), (e), and (f). (4)-(7): Four variants of hybrid quicksort using the ideas of (a)-(g), i.e., hyb1, hyb2, hyb3, and hyb4, with ($\alpha$, $\beta$) = (16, 16), (32, 32), (32, 64), (16, 32), respectively. The Java code of the hybrid quicksort is given in Appendix B. The BSD qsort is excluded here because it performs poorly on the $k$-sharp teeth instances when $k \geq 5$.

The performances of all methods on randomized inputs (the first five classes) are about the same, with the GNU qsort being the worst performer. The high cost on the class of random 256-lists is due to an expensive comparison function and high number of cache misses. Comparing to mergesort, quicksort as a group is slower on non-randomized inputs. However, quicksort's performances on these inputs are no worse than randomized inputs. In fact, the hybrid quicksort is the best performer over all the 12 classes of inputs. If we have to select one quicksort, then hyb2 ($\alpha = \beta = 32$) is the choice according to Table 3. The GNU qsort and Bentley and McIlroy's qsort [Bentley and McIlroy, 1993] are about 23% slower than hyb2. The use of idea (f) is indispensible for this performance as it helps quicksort to save time on nearly sorted inputs.

Sarwar *et al*. [Sarwar *et al*. 1996] once conducted an extensive experimentation of quicksort and tried various methods for selecting a pivot (from 1, 3, 5, 9, and 17 elements). They concluded that quicksort performs best when the pivot is the median of three elements and there is no need to call insertion sort. Our experiment shows that their claims are true only for randomized inputs. If we let $\beta = \infty$, quicksort will always pick the median of three elements as pivot, and this version of quicksort exhibits the O($n^2$) worst-case behavior for 2-equal teeth or 2-even teeth, which are called the organ-pipe shape inputs. The idea of (c), i.e., use the pseudo-median of 9 elements as pivot, essentially prevents the worse-case from happening. And the idea (a) helps to reduce the overhead of (c) when the array size is small. When $\alpha = \beta$, (b) is excluded and (c) becomes the only strategy for selecting a pivot. Our experiment shows that as long as $16 \leq \alpha \leq \beta \leq 100$, the performance of quicksort has little difference on random inputs.

David Musser's introsort [Musser, D.R., 1997] is an attempt to make quicksort adaptive to nearly sorted inputs by using heapsort when the recursive calls of quicksort go too deep ($\geq 2\log(n)$). The introsort in Table 3 calls insert sort when the subarray has no more than 32 elements and uses the median of three elements as pivot. For randomized inputs, it works as well as the simple 2-way quicksort when the median of three elements is used as pivot. When the elements have large sizes, its performance goes slightly down as heapsort is known to be poor on memory locality. However, for the classes of $k$-equal teeth, $k$-even teeth, $k$-sharp teeth, and $k$-exchanges, introsort is at least six times slower than the best performer. Our experiment shows that introsort performs poorly on nearly sorted inputs (82% slower) and is about 11% slower than the best quicksort (Introsort is about 10% faster than Bentley and McIlory's qsort). We have tested several versions of introsort and found that picking the pseudo median of nine elements as pivot and testing the sortedness of the input as a preprocessing step can speed up its performance on nearly sorted inputs. In short, mixing quicksort with heapsort is not a good way to handle nearly sorted inputs.

## 5.   COUNT NUMBERS OF COMPARISONS

In the previous sections, we used the running time as the metric to select the best performer. In the C library qsort function, the order and how to compare elements is defined by a comparison function which the caller must pass to qsort(). Since we control the comparison function and we use the interface of qsort() as a model for every sorting method, we can also count how many comparisons an implementation performs, which is also a pretty good metric.

Table 4: Average number of comparisons per element ($n$ = 2,000,000)

| cmp per item | 3-way | 2-way | hyb2 | hyb4 | B&M | qsort | mer4 | mer5 | mer6 | Tim |
|---|---|---|---|---|---|---|---|---|---|---|
| random 16-list | 22.51 | 24.62 | 24.47 | 23.47 | 22.57 | 19.68 | 21.62 | 20.77 | 21.81 | 19.20 |
| random 64-list | 22.74 | 24.73 | 24.70 | 23.41 | 22.08 | 19.67 | 21.62 | 21.09 | 22.22 | 19.64 |
| random 256-list | 22.92 | 24.83 | 24.80 | 23.50 | 22.07 | 19.67 | 21.62 | 21.04 | 22.17 | 19.64 |
| random double | 22.79 | 24.83 | 24.79 | 23.51 | 22.08 | 19.67 | 21.62 | 21.09 | 22.16 | 19.64 |
| random integer | 22.95 | 24.80 | 24.77 | 23.48 | 22.13 | 19.67 | 21.62 | 21.03 | 22.15 | 19.64 |
| k-limited | 13.65 | 16.64 | 14.63 | 14.06 | 13.30 | 18.89 | 15.89 | 15.23 | 16.17 | 14.19 |
| k-equal teeth | 19.24 | 22.86 | 19.68 | 19.24 | 20.59 | 11.23 | 4.02 | 4.02 | 4.02 | 4.02 |
| k-even teeth | 18.78 | 22.51 | 19.36 | 18.91 | 20.44 | 11.31 | 4.67 | 4.05 | 4.49 | 4.16 |
| k-sharp teeth | 6.65 | 6.39 | 6.46 | 6.58 | 19.90 | 9.32 | 1.33 | 0.89 | 1.33 | 0.89 |
| k-shuffled teeth | 19.07 | 20.54 | 20.48 | 19.49 | 21.61 | 16.57 | 8.52 | 7.18 | 8.43 | 6.71 |
| k-distance | 20.26 | 20.62 | 20.53 | 20.85 | 20.63 | 11.72 | 5.40 | 5.31 | 5.35 | 4.99 |
| k-exchange | 6.07 | 6.07 | 6.07 | 6.07 | 19.69 | 12.36 | 1.00 | 1.00 | 1.00 | 1.00 |
| Average cmp | 17.64 | 19.70 | 18.56 | 17.96 | 20.09 | 15.81 | 12.41 | 11.89 | 12.61 | 11.14 |

In Table 4, we give the average number of comparisons per item for the inputs in Table 3 ($n$ = 2,000,000) for the selected 10 sorting methods, the first five from quicksort, and the second five from mergesort. That is, each number is the average of 100 numbers of comparisons per item for randomized arrays. To compute the average number of comparisons for each $k$-class, we chose 9 values for $k$: The value of $k$ is $2^i$, where $i$ = 0, 1, …, 8. For each class, we took the average number of comparisons from 10 executions and then took the average for the whole class. The use of binary search in insertion sort can reduce the number of comparisons by about 10% (not shown here) for those methods using insertion sort.

For $n = 2,000,000$, hyb4 makes on average $19n$ comparisons for $k$-equal teeth and $k$-even teeth arrays, still lower than $23n$ comparisons for randomized lists (and much lower than $38n$ of heapsort for $k$-equal teeth and $k$-even teeth arrays). The idea of testing whether an array is already sorted is used in the first four quicksort methods. However, Table 4 does not show any significant increases in the number of comparisons. In fact, the small values for the $k$-sharp teeth and $k$-exchange classes tell that this idea is very useful.

Table 5: Value of $s$ in the number of comparisons function $f(n) = sn\log(n) + tn$

| value of s | 3-way | 2-way | hyb2 | hyb4 | B&M | qsort | mer4 | mer5 | mer6 | Tim |
|---|---|---|---|---|---|---|---|---|---|---|
| random 16-list | 1.03 | 1.06 | 1.01 | 1.04 | 1.10 | 1.00 | 1.13 | 1.02 | 1.01 | 1.02 |
| random 64-list | 1.14 | 1.11 | 1.11 | 1.11 | 1.13 | 1.01 | 1.13 | 1.02 | 1.01 | 1.02 |
| random 256-list | 1.09 | 1.07 | 1.07 | 1.07 | 1.07 | 1.01 | 1.13 | 1.03 | 1.05 | 1.02 |
| random double | 1.07 | 1.02 | 1.01 | 1.02 | 1.09 | 1.01 | 1.13 | 1.01 | 1.02 | 1.02 |
| random integer | 1.06 | 1.12 | 1.12 | 1.12 | 1.06 | 1.01 | 1.13 | 1.04 | 1.05 | 1.02 |
| k-limited | 0.46 | 0.50 | 0.46 | 0.48 | 0.48 | 0.96 | 0.54 | 0.49 | 0.50 | 0.51 |
| k-equal teeth | 1.24 | 1.27 | 1.23 | 1.23 | 1.31 | 0.41 | 0.01 | 0.01 | 0.01 | 0.00 |
| k-even teeth | 1.22 | 1.21 | 1.27 | 1.27 | 1.31 | 0.45 | 0.00 | 0.00 | 0.00 | 0.00 |
| k-sharp teeth | 0.04 | 0.03 | 0.03 | 0.02 | 0.98 | 0.45 | 0.00 | 0.00 | 0.00 | 0.00 |
| k-shuffled teeth | 0.79 | 0.77 | 0.78 | 0.79 | 1.07 | 0.84 | 0.00 | 0.00 | 0.00 | 0.00 |
| k-distance | 1.11 | 1.10 | 1.10 | 1.08 | 1.02 | 0.49 | 0.00 | 0.00 | 0.00 | 0.00 |
| k-exchange | 0.02 | 0.02 | 0.02 | 0.02 | 1.06 | 0.45 | 0.00 | 0.00 | 0.00 | 0.00 |

Table 6: Value of $t$ in the number of comparisons function $f(n) = sn\log(n) + tn$

| value of t | 3-way | 2-way | hyb2 | hyb4 | B&M | qsort | mer4 | mer5 | mer6 | Tim |
|---|---|---|---|---|---|---|---|---|---|---|
| random 16-list | -1.03 | 1.36 | 1.31 | 0.56 | -0.59 | -1.34 | -1.53 | -0.43 | 0.16 | -1.80 |
| random 64-list | -1.09 | 1.47 | 1.44 | 0.13 | -1.53 | -1.40 | -1.98 | -0.38 | 1.08 | -1.77 |
| random 256-list | 0.12 | 2.40 | 2.38 | 1.09 | -0.34 | -1.39 | -1.97 | -0.53 | 0.22 | -1.77 |
| random double | 0.47 | 3.47 | 3.69 | 2.16 | -0.76 | -1.38 | -1.94 | -0.11 | 0.74 | -1.81 |
| random integer | 0.83 | 1.33 | 1.40 | 0.01 | -0.15 | -1.39 | -1.95 | -0.80 | 0.13 | -1.79 |
| k-limited | 3.93 | 6.27 | 4.93 | 4.00 | 3.29 | -1.13 | 4.57 | 5.05 | 5.66 | 3.49 |
| k-equal teeth | -6.66 | -3.77 | -6.03 | -6.46 | -6.93 | 2.72 | 3.80 | 3.80 | 3.80 | 4.02 |
| k-even teeth | -6.85 | -2.83 | -7.33 | -7.77 | -7.06 | 1.86 | 4.67 | 4.05 | 4.49 | 4.16 |
| k-sharp teeth | 5.73 | 5.70 | 5.75 | 6.90 | -0.55 | -0.09 | 3.36 | 3.89 | 3.35 | 4.91 |
| k-shuffled teeth | 2.53 | 4.32 | 4.17 | 3.05 | -0.71 | -0.97 | 8.48 | 5.43 | 5.55 | 6.68 |
| k-distance | -2.91 | -2.46 | -2.47 | -1.67 | -0.68 | 1.42 | 5.73 | 5.32 | 5.36 | 5.53 |
| k-exchange | 5.65 | 5.65 | 5.65 | 5.65 | -2.41 | 2.92 | 1.05 | 1.05 | 1.05 | 1.05 |

As an approximation, we may assume that the number of comparisons is defined by a function $f(n) = sn\log(n) + tn$ on an input array of $n$ elements, where $s$ and $t$ are constants. Let $g(n) = f(n)/n = s\log(n) + t$, be the number of comparisons per item, which happens to be a line on $\log(n)$. To determine the values of $s$ and $t$, we run our implementations on the 12 classes of inputs of various sizes: there are five values for $n = j \cdot 10^6$, where $j = 1$ to 5. We then used linear regression to compute the approximate values of $s$ and $t$ (each line is computed over 5 points), which are given in Tables 5 and 6.

For randomized inputs, the values of $s$ for the 10 methods are close to 1, indicating these are very good sorting methods. For instance, for hyb4 over random doubles, the number of comparisons is $f(n) = 1.02n\log(n) + 2.16n$. These numbers also indicate that the relative strength of these method tend to be the same as comparison functions of different costs are used.

For the classes of $k$-equal teeth and $k$-even teeth arrays, the $s$ values of quicksort are slightly larger than 1, indicating apparently these are hard problems for quicksort.

For the last six classes, with the exception of the GNU qsort, the $s$ values of the mergesort methods are (almost) zero. That is, these methods take almost linear time to sort these six classes. That explains why these methods are so fast on these inputs.

## 6. CONCLUSIONS

Through extensive experiments, we have achieved our goals for this project.

- We have suggested 12 classes of inputs which exhibit different degree of presortedness and can be easily generated by a program. Using them, we can identify the best sorting methods and test any other method against them. For instance, we have tested a dozen of series for shellsort [Knuth, D.E. 1997] and found that the best shellsort is still 180% slower than the best mergesort or quicksort over the 12 classes of inputs. The biggest disappointment is heapsort, which has the optimal complexity of $O(n\log n)$, but performs poorly comparing to quicksort and mergesort. For the bottom-up heapsort [Williams, J. W. J. 1964], which is claimed to be much faster quicksort [Wegener, I. 1993], it is about three times slower over the 12 classes of inputs and 7 times slower over the eight classes of nearly sorted inputs. Smoothsort [Dijkstra, E. W. 1982], a variant of heapsort, runs about two times slower than quicksort. Splaysort [Moffat *et al*. 1996] is also much slower than mergesort and quicksort on nearly sorted inputs. All the sorting methods mentioned above and in the previous sections are implemented in C and compiled into a single executable called *sortrace*. The 12 inputs can also be generated from the same program. That is the software platform we plan to extend in the future, as we will add more types of inputs. The C code of *sortrace* is available as a supplement of this article.

- For sorting methods based on natural mergesort [McIlroy, P. 1993], we have identified three techniques of timsort [Peters, T., 2002] as important: (1) reversing revered runs; (2) the galloping merge, and (3) management of runs to decide which and when to merge two runs. The weakness of neatsort [La Rocca *et al*. 2014] is perhaps due to the missing of (2). We proposed a new way to manage runs (in mer5 and mer6 using the half-down list) which has better performance than timsort for nearly sorted inputs. The use of insertion sort is helpful as evident by the slight gain of mer6 over mer5. The use of binary search in insertion sort has high impact on the number of comparisons but little impact on running time. The adjustment of the minimal run length is easy to implement but does not have any significant impact on performance.

- For sorting methods based on quicksort, we confirmed that Bentley and McIlroy's method of choosing a pseudo-median over 9 elements as pivot is crucial to avoid quicksort's worst case complexity of $O(n^2)$. We proposed two new ideas to improve quicksort: (1) testing sortedness as a preprocessing step in each recursive call; (2) a hybrid quicksort combing the 2-way quicksort with the 3-way quicksort. The idea of (1) is to make quicksort adaptive to nearly sorted inputs. The idea of (2) is to avoid the overhead of 3-way splitting when there are few identical elements. As a result, our quicksort is the overall champion for the 12 classes of inputs. For the eight contrived classes of inputs, our quicksort also beats the GNU qsort and heapsort. For the textbook version of the quicksort algorithm [Cormen *et al.*, 2009], the worst-case time complexity comes when the input is a $k$-even teeth or $k$-equal teeth. This property was regarded as a significant drawback of quicksort. Our experiment shows that using the proposed techniques in this article, we can not only avoid the worst-case complexity completely but also make quicksort adaptive to nearly sorted inputs. Apparently, the C library should not shun quicksort.

We have tried to avoid theoretic analysis and technical details in this article. As further research, we are interested in a formal analysis of the adaptiveness of quicksort when sortedness testing is used as a preprocessing step. Apparently such testing uses more comparisons. However, the experiment shows no significant increase in the number of comparisons. For instance, for hyb4 on $n$ randomized doubles, the number of comparisons is $f(n) = 1.02n\log(n) + 2.16n$, while Bentley and McIlroy's qsort uses $f(n) = 1.09n\log(n) - 0.76n$. Moreover, we would like to see how the complexity of mergsort and quicksort is related to the metric *Mono(A)*.

One additional advantage of mergesort over quicksort is that mergesort is stable. In general, mergesort takes $O(n)$ more memory than quicksort as required by merging. In our experiments, we see that both methods slow down on long lists of heavy items. Timsort works better than our mergesort on such case, but is still slower than quicksort. Our experiment has been run on a linux machine with an Intel(R) Xeon(R) CPU E5-2667 (20M Cache and 3.20GHz) and 64GB memory. Further experiments are needed to test the limits of both methods and search for better techniques to deal with memory shortage. Further investigation is needed to study memory access patterns of sorting algorithms. It is known that both mergesort and quicksort enjoy much better memory locality than shellsort and heapsort, and this may explain why mergesort and quicksort perform better than shellsort and heapsort in our experiments. The optimization techniques related to memory locality, such as branch prediction, can affect the performance of sorting algorithms. Besides cache size, the entire memory system may affect the performances of sorting algorithms. We would like to test our program on different machines and investigate the impact of cache memory. In addition to different laptop/desktop configurations, it would be interesting to see performance of these sorting algorithms for mobile platforms such as the android platform.

We observe that calling the comparison function in C takes significant more time than the same sorting method without calling the comparison function (such as using micros or inline functions in C++/Java). Since our goal is to address algorithmic issues of sorting methods, we are not concerned about this slowness, as every method in our C program is treated the same way, so it is fair. It would be interesting to see how the methods implemented in our platform perform when language-dependent optimization techniques are used or in other programming language environments, such as in C++ STL (Standard Template Library), where templates can be used to avoid function calls.

Regarding the question raised in the beginning of this article, "if you are asked to recommend a sorting algorithm for an unknown problem, what is your recommendation?" What will be your answer after reading this article? Our answer is quicksort, especially our hybrid quicksort, if stable sorting is not required. Only when we know that stable sorting is required, or the input contains some kind of presortedness and the memory is not a concern, then mergesort, either timsort or ours.

**ACKNOLEDGEMENTS**

Special thanks go to Marcello La Rocca who converted their neatsort into C and gave it to us. The C codes of the BSD qsort, timsort, smoothsort, bottom-up heapsort, and splaysort are from the following websites:

> http://opensource.apple.com//source/xnu/xnu-1456.1.26/bsd/kern/qsort.c
> https://github.com/patperry/timsort,
> https://en.wikibooks.org/wiki/Algorithm_Implementation/Sorting/Smoothsort,
> https://github.com/Maxime2/heapsort/blob/master/dps_bottom_up_heapsort.c,
> http://people.eng.unimelb.edu.au/ammoffat/splaysort.c.

We thank the authors of these codes for making them available.

**APPENDIX A. The C code of our mergesort**

// merge(arr, left, middle, right) merges arr[left, middle-1] and arr[middle, right-1] into arr[left, right-1]
// reverse(arr, left, right) reverse the subarray arr[left, right].

```
void mergesort6 (WORD *arr, size_t n, size_t es, int (*cmp)(const void *, const void *)) {
   int *run, next, j, s, t, runnum, start;
   run = (int *) malloc(64*sizeof(int));  // size of run is log(n) – 64 is more than enough.

   runnum = run[0] = start = 0; next=1;
   while (next < n) {
      if (cmp(arr+next-1, arr+next)>0) {   // i.e., if (arr[next-1] > arr[next])

        if (start+minrun > next) {     // make sure each run has at least minrun elements
            j = next+1; s = 0;  // look for reversed run and reverse equal elements
            while (j < n && ((t=cmp(arr+j-1, arr+j))>=0)) {
              if (t == 0) s++;
              else if (s > 0) { reverse(arr, j-1-s, j-1);  s = 0;  }   // i.e., reverse arr[j-1-s .. j-1].
              j++;
            }
            if (next > start+1) {
                reverse(arr, next, j-1);           // reverse from next to j-1, then merge two parts.
                merge(arr, start, next, j);
            } else  { reverse(arr, start, j-1); }  // reverse from start to j-1
            next = j;

            if (start+minrun > next)  {      // still too short, use insertion sort to extend the run
               j = (start+minrun<n)? start+minrun : n;
               isort(arr, start, next, j);      // insertion sort arr[start..j-1] where arr[next..next-1] is sorted.
               next = j;
            } else { // of (start+minrun > next)
               // record a new run
               run[++runnum] = start = next++;

               // merge two runs if the current  run > 1/2 of its predecessor
               while (runnum > 1 && (run[runnum]-run[runnum-1]) > (run[runnum-1]-run[runnum-2])>>1) {
                   j = run[runnum-2];
                   merge(arr, j, run[runnum-1], run[runnum]);
                   run[runnum-1] = run[runnum];
                   runnum--;
               }
            }
         } else next++;
     }

     // merge all runs into one, from last to first.
     while (runnum > 0) {
         j = run[runnum-1];
         merge(arr, j, run[runnum--], n);
     }
     free(run);
}
```

**APPENDIX B. The java code of hybrid quicksort**

To better present our ideas, we give below the Java code of our algorithm, while the experiment is conducted using a C implementation.

```java
public void qsort_hyb(Item[] arr, int low, int high) {
    Item v;
    int i, j, r, c=0, n=high-low+1;

    if (n < alpha) { isort(arr, low, high); return; }
    if (is_sorted(arr, low, high)) return;
    if (n < beta) v = pick_pivot3(arr, low, high);         // v is the pivot
    else v = pick_pivot9(arr, low, high);                  // v is the pivot

    i = low; j = high;
    while (true) {
        while ((r=v.compareTo(arr[i])) < 0) i++;  if (r==0) c++;
        while ((r=v.compareTo(arr[j])) > 0) j--;  if (r==0) c++;
        if (i >= j) break;
        exchange(arr, i++, j--);
    }

    if (c > 2) {
       if (low < j) qsort_3way(arr, low, j);
       if (i < high) qsort_3way(arr, i, high);
    } else {
       if (low < j) qsort_hyb(arr, low, j);
       if (i < high) qsort_hyb(arr, i, high);
    }
 }
```

The following methods are used in the method qsort_hyp:
```java
        // insert sort on subarray arr[low..high]:
        public void isort(Item[] arr, int low, int high) { }
        // return true iff arr[low..high] is sorted:
        public boolean is_sorted(Item[] arr, int low, int high) { }
        // return the median of arr[low], arr[(low+high)/2], and arr[high]:
        public Item pick_pivot3(Item[] arr, int low, int high) { }
        // return the pseudo-median of nine elements:
        public Item pick_pivot9(Item[] arr, int low, int high) { }
        // exchange elements at positions i and j in arr:
        public void exchange(Item[] arr, int i, int j) { }
        // an implementation of 3-way quicksort:
        public void qsort_3way(Item[] arr, int low, int high) { }
```

Note that integers alpha and beta are cutoff values for using isort, pick_pivot3, or pick_pivot9. Besides adding the test is_sorted and selecting better pivots, this version of quicksort is different the textbook version of quicksort [Cormen *et al.*, 2009] in that the position of the pivot element is not identified and there is no attempt to move the pivot element inbetween the two subarrays after the splitting.